\documentclass{epl}
\usepackage{graphicx}
\usepackage{bm}

\title{Multifractal clustering of passive tracers on a surface flow}
\author{G. Boffetta\inst{1} , J. Davoudi\inst{2},
        F. De Lillo\inst{3}}
\institute{
  \inst{1} Dipartimento di Fisica Generale and INFN,
           Universit\`a degli Studi di Torino,
           Via Pietro Giuria 1, 10125, Torino, Italy \\
           and ISAC-CNR, Sezione di Torino, Italy \\
  \inst{2}  Fachbereich Physik, Philipps-Universit\"at, Renthof 6,
            D-35032 Marburg, Germany \\
  \inst{3} INLN-CNRS, 1361 route des Lucioles, Sophia Antipolis, 06560 Valbonne, France.
}

\begin{document}
\def\be{\begin{equation}}
\def\ee{\end{equation}}
\def\bea{\begin{eqnarray}}
\def\eea{\end{eqnarray}}

\maketitle
\begin{abstract}

We study the anomalous scaling of the mass density measure of Lagrangian 
tracers in a compressible flow realized on the free surface on top of a 
three dimensional flow.
The full two dimensional probability distribution of local stretching rates 
is measured. The intermittency exponents which quantify the fluctuations of 
the mass measure of tracers at small scales are calculated from the large 
deviation form of stretching rate fluctuations. 
The results indicate the existence of a critical exponent $n_c \simeq 0.86$
above which exponents saturate, in agreement with what has been predicted 
by an analytically solvable model.
Direct evaluation of the multi-fractal dimensions by reconstructing the 
coarse-grained particle density supports the results for low order moments.   

\end{abstract}

Advection of homogeneous distribution of passive particles in a 
compressible flow generically results in clusters of particles' 
concentration \cite{FGV01,BFF01,ott}.
The effects of compressibility in three-dimensional flows are 
typically relevant only at Mach numbers near or larger than $1$
when density inhomogeneity reacts back on the fluid velocity. 
However there are simple physical systems in which the phenomenology 
of passive particle advection in compressible flows is applicable. 
Effective flows of inertial particles in incompressible flows at
small Stokes numbers \cite{Bec03,KS97,EKLRS02} and flows of surface 
suspensions \cite{GCVES01,NJP,BDES04} are among such instances.

Consider the distribution of particles advected by a
velocity field ${\bf u}({\bf x},t)$ assumed homogeneous and
stationary in time. The evolution of particle density 
$\rho({\bf x},t)$ in the limit of vanishing diffusivity is given by
\be
\frac{\partial \rho }{\partial t}+\nabla \cdot (\rho {\bf u})=0
\label{eq:1}
\ee
If ${\bf \nabla} \cdot {\bf u} \ne 0$ the flow is compressible and
particle trajectories asymptotically converge to a non-uniform distribution.
Compressibility can be characterized by the
dimensionless ratio
${\cal C} = {\langle \left(\partial_i u_i \right)^2 \rangle
\over \langle \left(\partial_i u_j \right)^2 \rangle }$
which assumes values in $0\le {\cal C} \le 1$, the limiting values 
corresponding to an incompressible and a potential flow respectively.
The particle density depends on the particular realization of the 
velocity field and for a stationary random flow the density is expected 
to converge to a statistically stationary state $\rho_*$.

A suitable way to characterize the statistics of the density field 
on the attractor is to measure the 
{\it Lagrangian} moments of the coarse-grained mass in a small 
region of size $r$ as:
\bea
\langle m_r^n \rangle= 
\overline{\int \rho_*({\bf x})m_r^n({\bf x}) d{\bf x}}
\label{eq:2}
\eea
where $m_r({\bf x})=\int_{{\cal B}_r({\bf x})}\rho_*({\bf x'})d{\bf x'}$ is 
the coarse-grained mass in a ball of radius $r$ centered at ${\bf x}$. 
The weight $\rho_*$ in (\ref{eq:2}) restricts the average to balls 
centered on the attractor.
Note that the coarse grained mass $m_r({\bf x})$ depends on the specific 
velocity configuration within a given ensemble of flow realizations:
the $\overline {(\cdots)}$ means the averaging with respect to the given 
velocity statistics.

When $r\to 0$ and under generic assumptions for the random flow $\langle 
m_r^n \rangle \sim r^{\xi_n}$ \cite{ott,HP83}.
A non-linear dependence of the set of scaling exponents $\xi_n$ 
on the order $n$ implies an {\it intermittent} distribution of 
mass fluctuations at small scales.
Recently, a phenomenological connection between the probability 
distribution of small scale velocity stretching rates and scaling 
exponents of coarse grained mass moments has been obtained \cite{BGH04}.
More precisely the set of exponents $\xi_n$ are bridged to the Cramer 
function of the stretching rates of the underlying flow. 
It is shown that $\xi_n$ saturates at large $n$. The saturation is a 
footprint of the very violent events which spoil the scaling in the 
higher mass moments.

Here we study the clustering of passive tracers which move 
on the two-dimensional free slip surface on the top of a three-dimensional
incompressible fluid \cite{BDES04,NJP}. 
Our main result is that an intermittent distribution of passive tracers,
with saturation of high-order scaling exponents, 
is observed also in this realistic, time correlated flow.

When the flow is spatially smooth the evolution of trajectories is a 
combination of random stretching and contracting in time. 
The individual tracers are advected by the same realization of the 
flow thus the evolution of the separation between two 
Lagrangian particles is given by 
${\bf X}(t,{\bf x})={\bf W}(t,{\bf x}) {\bf X}(0,{\bf x})$ where 
${\bf X}(0,{\bf x})$ is the initial separation. 
The linear random stretching and contracting is encoded in the 
positive eigenvalues of the symmetric matrix $W^{\dagger}(t,x)W(t,x)$:
$e^{2 t \sigma_1},\cdots,e^{2\ t \sigma_d}$.
The {\it forward stretching rates} $\sigma_i(t)$ are arranged
in non-decreasing order and asymptotically converge to the 
{\it Lyapunov } exponents at large times, i.e. $\lambda_i=\lim_{t 
\to \infty}\sigma_i$ \cite{ott}.
For generic cases when the spectrum is non-degenerate the local 
stretching rates satisfy an asymptotic probability distribution function 
(PDF) with the following large deviation shape \cite{BF99},
\be 
P(\sigma_1,\cdots,\sigma_d;t) \sim e^{-t H(\sigma_1,\cdots,\sigma_d)},
\ee
where the Cramer function $H(\sigma_1,\cdots,\sigma_d) \geq 0$ is convex and takes its 
minimum in 
$H(\lambda_i)=0$.

For completeness, here we briefly sketch the outline of the theoretical 
arguments first appeared in \cite{BGH04}.
In a two-dimensional compressible flow, the Lagrangian evolution
correspond to a dissipative dynamics with $\sigma_1 > 0$ and $\sigma_2 <0$
and with $\sum_i \sigma_i \le 0$.
Along the unstable direction the density is smooth while in the stable 
direction the density configuration is fractal.
After a transient regime the density is expected to reach to 
statistical stationary state.
We will take formally the initial time $t_0 \to -\infty$ thus we will
assume that at $t=0$ the density $\rho_*$ is already stationary.
At time $t>0$ one looks at the fluctuation of the 
coarse-grained mass generated by $\rho_*$ according to (\ref{eq:2}).
Consider a rectangular box of size $r$ with sides $r_1$ and $r_2$ along the 
orthogonal unstable and stable directions. 
According to eq.(\ref{eq:2}) $\langle 
m_{r_1,r_2}^n \rangle \sim r_1^{\xi_n^{(1)}} r_2^{\xi_n^{(2)}}$ where 
$\xi_n=\xi_n^{(1)}+\xi_n^{(2)}$.
Since the measure along the unstable direction $r_1$ is regular one 
gets $\xi_n^{(1)}=\min(\xi_n,n)$.
For $\xi_n \geq n$ it results to $\xi_n^{(1)}=n$ and therefore 
$\langle m_{r_1,r_2}^n \rangle \sim r_1^n r_2^{\xi_n-n}$. 
In the case of $\xi_n < n$ it reads 
$\langle m_{r_1,r_2}^n \rangle \sim r_1^{\xi_n} r_2^0$.

The pre-image of the box at time $t=0$ is a stretched box as 
$r_1$ shrinks to $r_1e^{-t\sigma_1}$ and $r_2$ increases to $r_2^{-t\sigma_2}$.
Due to conservation of the mass and stationarity the mass moments 
in the pre-image box at time $t=0$ are the same, i.e. 
\be
\langle m_{r_1,r_2}^n \rangle \sim 
\langle m_{r_1e^{-t\sigma_1},r_2e^{-t\sigma_2}}^n\rangle.
\label{eq:4}
\ee
According to the definition in (\ref{eq:2}) the average on the right 
hand side is considered over the distribution of the forward 
stretching rate with respect to invariant measure $\rho_*$. 
Note that the result of averaging over forward and backward 
distributions with respect to invariant measure 
are the same \cite{private}. 
At large time $t$ one resorts on the saddle point argument whereby two distinct situations 
emerges from eq.(\ref{eq:4}) that is,
\begin{eqnarray}
&&\min_{\sigma_1,\sigma_2}\{ H(\sigma_1,\sigma_2)+n\sigma_1+
(\xi_n-n)\sigma_2\}=0 \quad n\le \xi_n,
\label{eq:5}
\\[5pt]
&&\min_{\sigma_1,\sigma_2}\{H(\sigma_1,\sigma_2)+\xi_n\sigma_1\}=0 
\quad n \ge \xi_n .
\label{eq:6}
\end{eqnarray}
The above relations connect the scaling exponents $\xi_n$ to the Cramer 
function $H(\sigma_1,\sigma_2)$.
For a synthetic Gaussian and white in time smooth flows (the Kraichnan
model of turbulence \cite{K94}) where the exact functional form of 
Cramer function is known, eqs. (\ref{eq:5}-\ref{eq:6}) lead to an analytic 
form for the scaling exponents \cite{BGH04}:

\begin{equation}
\xi_n = 
\left\{\begin{array}{cl}
 \frac{2n + \sqrt{(1+2{\cal C})^2 - 8{\cal C} n^2}}{1+2{\cal C}} -1 &
  \ \mbox{ if }\quad n \le n_{\rm c}\,,
\\[5pt]
 \xi_{\infty} &
  \ \mbox{ if }\quad n \ge n_{\rm c}
  \end{array}\right.
\label{eq:7}
\end{equation}
where the critical order for saturation is given by
\begin{equation} 
n_{\rm c}=
\left\{\begin{array}{cl}
 \frac{1}{2}\sqrt{1+\frac{1}{2{\cal C}}} &
 \ \mbox{ if }\quad 0 < {\cal C} \le \frac{1}{6} \,,
 \\[5pt] 
 \frac{2-4{\cal C}}{1+2{\cal C}} &
 \ \mbox{ if }\quad \frac{1}{6} \le {\cal C} < \frac{1}{2} \,.
 \end{array}\right.
\label{eq:8}
\end{equation}

In the general case, depending on the flow compressibility ${\cal C}$ 
two scenarios for saturation are thus conceivable.
In the first scenario the line $\xi_n=\xi_{\infty}$ from (\ref{eq:6}) 
intersects with the curve of $\xi_n(n)$ given by (\ref{eq:5}).   
In such a case $\xi_{\infty}=\xi_{n_c}=n_c$ and
the value of $(\sigma^*_1(n_c),\sigma_2^*(n_c))$ will pass through 
the line $\sigma_1=\sigma_2$. 
In the second case the line tangent to the maximum of 
the manifold $\xi_n(n)$ given by (\ref{eq:5}) identifies $\xi_{\infty}$.
In this later scenario $\xi_{n_c}\neq n_c$ and $n_c$ identifies with 
the solution of $\frac{d \xi_n}{d n}|_{n_c}=0$.
For the Kraichnan model the second case results in 
$\xi_{\infty}=\xi_{n_c}=2n_c-1$.

We have integrated the three-dimensional Navier-Stokes equation for an
incompressible flow by means of a standard pseudo-spectral code
at resolution $64^3$. Lateral boundary conditions for $x$ and $y$ directions
are periodic, while in the vertical $z$ direction we apply free-slip
boundary conditions, $u_z=0$ and $\partial_z u_x=\partial_z u_y=0$
\cite{GCVES01}.
When the flow is in statistically stationary conditions, $10^5$ passive
tracers are placed on the top free surface $z=0$. Their two dimensional
positions ${\bf x}(t)=(x(t),y(t))$ are advected by the two-dimensional
flow $(u_x,u_y)$ which is compressible since 
$\partial_x u_x + \partial_y u_y = - \partial_z u_z \neq 0$ 
\cite{NJP,BDES04}.
The compressibility of the surface flow was found to be 
${\cal C} \simeq 0.21$.

\begin{figure}[htb]
\centerline{
\includegraphics[scale=0.5,draft=false]{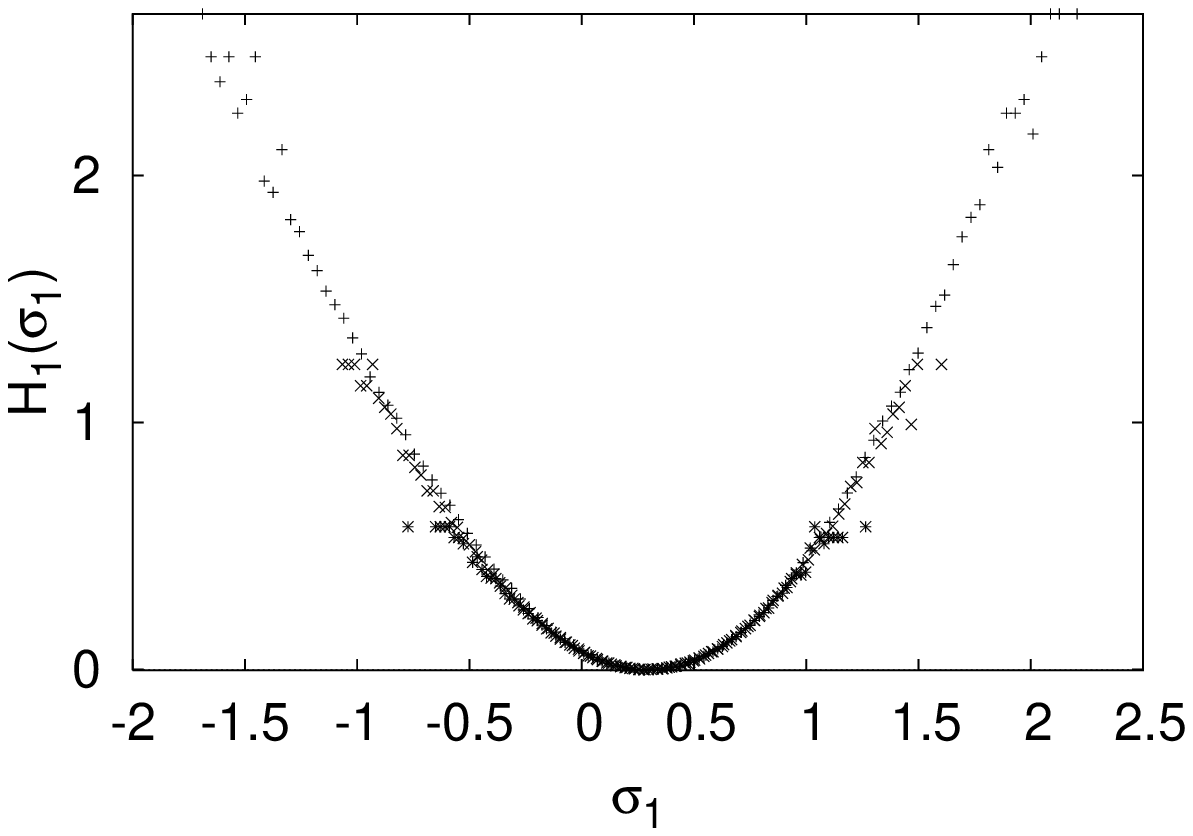}
\includegraphics[scale=0.5,draft=false]{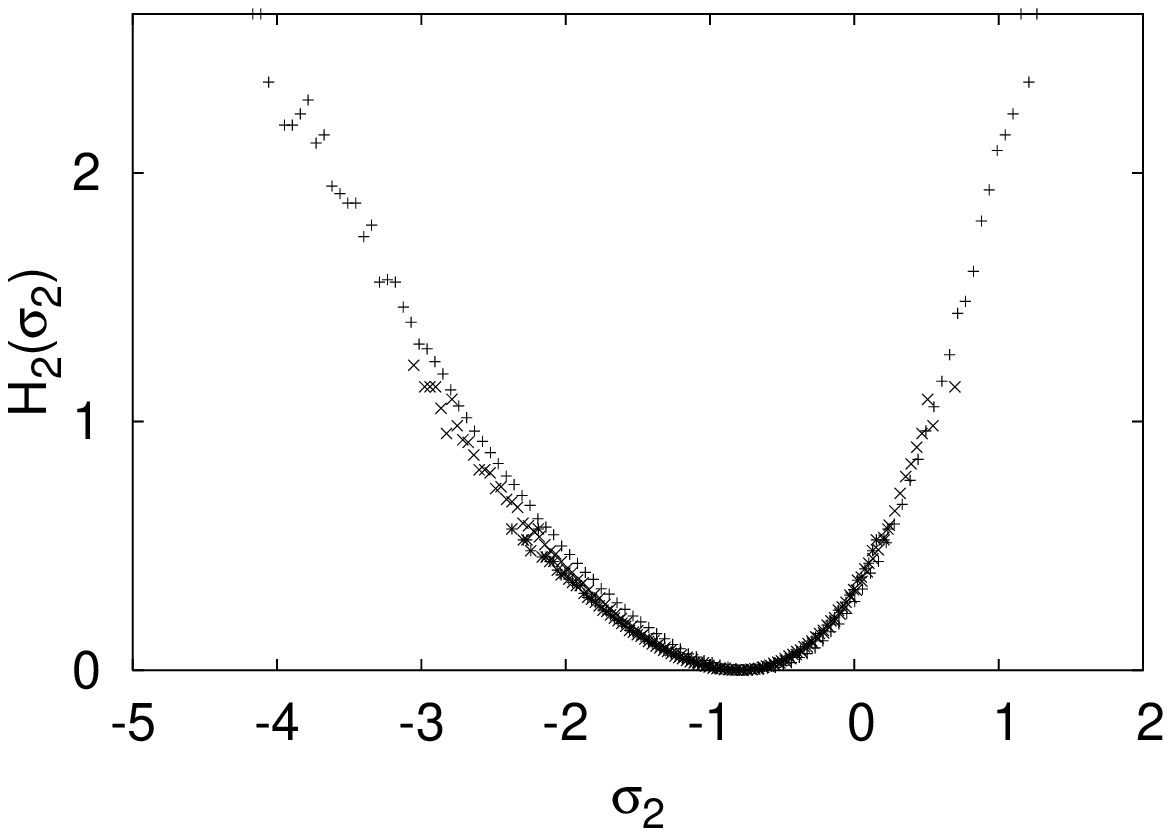}
}
\caption{Marginal Cramer functions $H_1(\sigma_1)$ (left) and $H_2(\sigma_2)$ 
(right) computed at three different times lags $\tau=\lambda_1^{-1}$, 
$\tau=2\lambda_1^{-1}$ and $\tau=4\lambda_1^{-1}$.
}
\label{fig1}
\end{figure}

In order to evaluate the stretching rates, we have integrated along each
trajectory a set of two orthogonal tangent vectors ${\bm z}^{(k)}$ 
($k=1,2$) evolving according to
\begin{equation}
{d \over dt} z^{(k)}_i = (\nabla_j u_i) z^{(k)}_j \quad (i,j=1,2)
\label{eq:9}
\end{equation}
The PDF of stretching rates at fixed time lag $\tau$,
$P(\sigma_1,\sigma_2;\tau)$, is then accumulated by computing
$\sigma_{k}=(1/\tau) \ln(|z^{(k)}(\tau)|/|z^{(k)}(0)|)$ for
different realizations along the trajectories.

Stretching rate statistics are accumulated up to
$t_{max}=15000 \lambda_1^{-1}$ which correspond to about 
$7000$ velocity correlation times $T \simeq 2.1 \lambda_1^{-1}$.
The fluctuations of each of the stretching rates is quantified by the   
convergence of their corresponding marginal Cramer functions $H_i(\sigma_i)$.
These are defined as
\be
P_i(\sigma_i;t)=\int P(\sigma_1,\sigma_2;t) d\sigma_{3-i} \sim 
e^{-t H_i(\sigma_i)} \,\,;\,\,{\tiny i=1,2},
\label{eq:10}
\ee

Fig.\ref{fig1} shows the two marginal Cramer functions obtained
from the marginal probability density functions of stretching rates
at three different time lags
$\tau=\lambda_1^{-1}$, $\tau=2\lambda_1^{-1}$ and $\tau=4\lambda_1^{-1}$.
The average values of stretching rates, i.e. the minimum of the 
marginal Cramer functions, converge to the Lyapunov exponents
$\lambda_1\simeq 0.29$, $\lambda_2 \simeq -0.86$ from which
one estimates the Lyapunov dimension \cite{ott} as 
$D_L=1+\frac{\lambda_1}{|\lambda_2|}\simeq 1.34$.
This value is larger to what obtained in a similar set of simulations
at higher Reynolds numbers \cite{BDES04}, but is consistent with the
fact that here the value of compressibility is smaller.

\begin{figure}[htb]
\centerline{
\includegraphics[scale=0.8,draft=false]{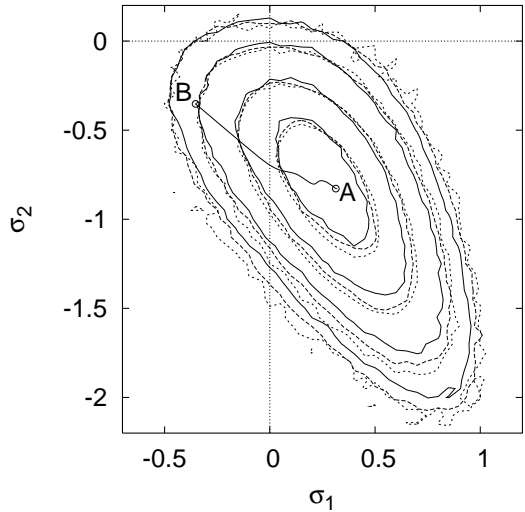}
}
\caption{
Two dimensional contour plots of the Cramer functions $H(\sigma_1,\sigma_2)$ 
computed at three different times $\tau=\lambda^{-1}$, $\tau=2\lambda^{-1}$
and $\tau=4\lambda^{-1}$. The geometrical locus of 
$(\sigma_1^*(n),\sigma_2^*(n))$ 
is shown from $n=0$ (A) to $n=n_c\simeq 0.86$ (B).}
\label{fig2}
\end{figure}

In Fig.(\ref{fig2}) we present the contour plot of the two
dimensional Cramer function $H(\sigma_1,\sigma_2)$  evaluated
at three different time lags.
The overlap of the contour corresponding to different time delays
is the indication of the convergence of the statistics.
We observe that the contours are rather far from an
elliptic shape, a direct indication of the deviations of
the probability density $P(\sigma_1,\sigma_2;\tau)$  
from a Gaussian distribution (as it is also evident from the
marginal distributions in Fig.~\ref{fig1}).

From the Cramer function $H(\sigma_1,\sigma_2)$ the set of
scaling exponents $\xi_n$ is obtained by numerical minimization
of relations (\ref{eq:5})-(\ref{eq:6}). For each order $n$ this
procedure determines a point $(\sigma_1^*(n),\sigma_2^*(n))$ in 
the stretching rate plane. The line in Fig.(\ref{fig2})     
represents the trajectory of these points starting from $n=0$
to $n=n_c\simeq 0.86$ where the point crosses the line
$\sigma_1=\sigma_2$ and the exponents saturates, for $n \ge n_c$,
to $\xi_n=n_c$, following the second
of the two scenarios described above.

\begin{figure}[htb]
\centerline{
\includegraphics[angle=0,scale=0.8,draft=false]{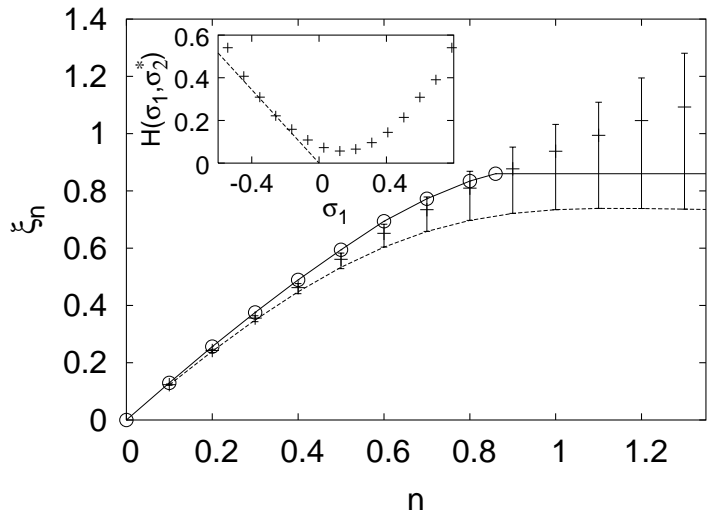}}
\caption{
Scaling exponents $\xi_n$ computed from the stretching-rate analysis
(connected circles). The horizontal line for $n>n_c\simeq 0.86$
represents the saturation value $\xi_\infty=n_c$. 
Scaling exponents obtained from a direct box-counting analysis
are shown after averaging $60$ sets of $200$ configurations each
($+$, error bars estimated on the fluctuations among different sets).
The lower line is the result of the direct computation of mass exponents 
for the particular subset where saturation is observed (see Fig.~\ref{fig4}).
}
\label{fig3}
\end{figure}

The set of scaling exponents obtained from the minimization of
Cramer function is shown in Fig.~\ref{fig3}.
Intermittency of mass distribution is evident from the non-linear
behavior of $\xi_{n}$.

The geometrical interpretation of the saturation predicted by 
(\ref{eq:6}) is that the intersection between the line $-\xi_{n_c} \sigma_1$ 
and the curve $H(\sigma_1,\sigma_2^*(n_c))$ is tangent at $n=n_c$.
In the inset of Fig.~\ref{fig3} we show that this is indeed reproduced
by our data and that the tangent point is well inside the
convergence range. This result indicates that the saturation observed
in Fig.~\ref{fig3} is a genuine phenomenon and it is not induced by 
finite size statistics.

We have compared the scaling exponents determined by the stretching rates
with those directly measured from the mass distribution.
To this aim, we stored $12000$ configurations of the $10^5$
particles at an interval $\Delta t=0.67 T$.
Particle density $\rho({\bm x})$ is reconstructed by distributing
the positions of the Lagrangian tracers on a regular grid of
resolution $512 \times 512$.
For each snapshot we directly measure the
coarse-grained mass and calculate the mass moments
$\langle m_r^n \rangle$ according to Eq.(\ref{eq:2}).
The moments are then averaged over the $12000$ snapshots
and the scaling exponents are estimated by a best fit.
The box-counting scaling exponents $\xi^{(BC)}_n$ are shown
in Fig.~\ref{fig3} up to order $n=1.3$.

As it is evident from the figure the moments are extremely fluctuating.
The low order exponents, for $n \leq n_c$, are very close to the
values obtained from minimization of Cramer function.
The saturation of the scaling exponents is corresponding to the rare 
events when large fraction of the particles accumulate at very small scales.
For $n>n_c$ the box counting exponents deviate from the saturation
line, but we have found that the distribution strongly fluctuates from
one realization to another. 

\begin{figure}[htb]
\centerline{
\includegraphics[angle=0,scale=0.8,draft=false]{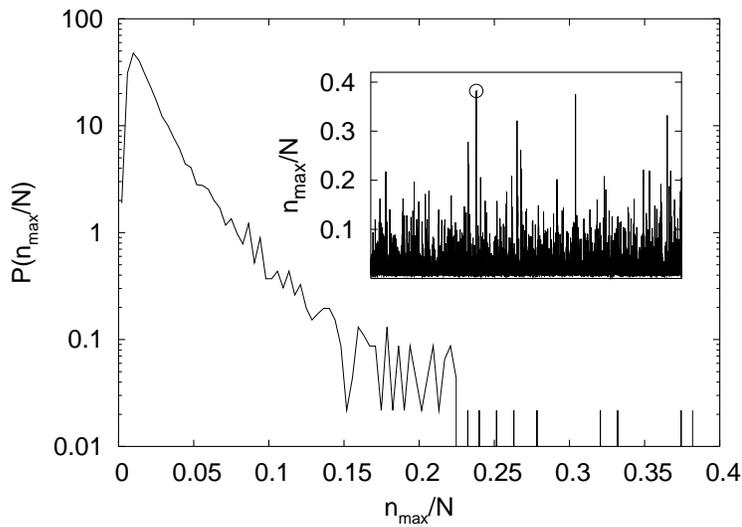}}
\caption{
The distribution of the maximum fraction of particles contained in a  
single bin of a $256\times 256$ grid. In the inset the time series of
$n_\mathrm{max}/N$ is shown. The circled point is the "singular" event which is
contained in the subset from which the dashed curve of Fig.~\ref{fig3} was
extracted.
}
\label{fig4}  
\end{figure}

In order to estimate the magnitude of these fluctuations, we have computed,
for each realization, the maximum fraction of particles in a single 
grid cell, i.e. $n_\mathrm{max}/N$. The resulting distribution
is shown in Fig~\ref{fig4} which shows a rather fat tail with 
events up to $n_\mathrm{max}/N \simeq 0.4$. 
This value of concentration has to be compared with the most probable 
value $n_\mathrm{max}/N \simeq 0.009$ (see Fig~\ref{fig4}).

These configurations of extreme concentration are responsible for the 
saturation of scaling exponents. 
Indeed, if we compute the box-counting exponents
$\xi^{(BC)}_n$ for one of these particular configuration, we observe
a saturation of the exponents at a value even smaller than 
$\xi_{\infty}$ (see Fig.~\ref{fig3}).
Although our data provides sufficient statistics to measure the joint 
distribution of local stretching rates the dynamical evolution of the 
rare events responsible for saturation of the coarse grained mass 
requires by far a longer sampling. This explains the large level of 
fluctuation for $\xi^{(BC)}_n\geq n$ in Fig.~\ref{fig3}. 

In summary we have obtained the two dimensional distribution of
the local stretching rate in a $2D$ surface flow with compressibility 
${\cal C}=0.21$.
The spectrum of the scaling exponents of the
intermittent distribution of the coarse grained mass at small scales has been  
numerically computed.
Our results indicate that the mass moments saturate at order $n_c \simeq 0.86$ 
in qualitative agreement with the prediction based on a Gaussian,
white in time model of compressible flow.

Direct measurement of the coarse grained mass density at small scales
confirms the results within the error bars up to order $n \simeq 1.3$.
The events of high mass concentration at small scales have been identified
by calculating the probability density of the maximum number of 
particles on grid resolution. 
The multi-fractal spectrum obtained from the coarse grained moments 
for those specific events is compatible with the existence of a
saturation exponent as predicted by the stretching rate analysis.

\acknowledgments
This work was supported by the Deutsche Forschungsgemeinschaft.
Numerical simulations have been performed at CINECA (INFM parallel
computing initiative).


\begin{thebibliography}{0}

\bibitem{BFF01}
\Name{Balkovsky A., Falkovich G. \and Fouxon A.}
\REVIEW{Phys. Rev. Lett.}{86}{2001}{2790}

\bibitem{FGV01}
\Name{Falkovich G., Gaw\c{e}dzki K. \and Vergassola M.}
\REVIEW{Rev. Mod. Phys.}{73}{2001}{913}.

\bibitem{ott}  
\Name{Ott E.}
\Book{Chaos in Dynamical Systems}
\Publ{Cambridge University Press}
\Year{1993}.



\bibitem{Bec03}
\Name{Bec J.}
\REVIEW{Phys. Fluids}{15}{2003}{L81}.

\bibitem{KS97}
\Name{Klyatskin V. I. \and Saichev A. I.}
\REVIEW{JETP}{84}{1997}{716}

\bibitem{EKLRS02}
\Name{Elperin T., Kleeroin N., L'vov V.S., Rogachevskii I. \and Sokoloff D.}
\REVIEW{Phys. Rev. E}{66}{2002}{036302}

\bibitem{GCVES01}
\Name{Goldburg W. I., Cressman J. R., V\"or\"os Z., Eckhardt B. 
\and Schumacher J.}
\REVIEW{Phys. Rev. E}{63}{2001}{065303}.


\bibitem{NJP}
\Name{Cressman J. R., Davoudi J., Goldburg W. I., \and Schumacher J}
\REVIEW{New J. Phys.}{6}{2004}{53}
 
\bibitem{BDES04}
\Name{Boffetta G., Davoudi J., Eckhardt B. \and Schumacher J.}
\REVIEW{Phys. Rev. Lett.}{93}{2004}{134501}

\bibitem{HP83}
\Name{Hentschel H.G.E. \and Procaccia I.}
\REVIEW{Physica D}{8}{1983}{435}




\bibitem{BGH04}
\Name{Bec J., Gaw\c{e}dzki K. \and Horvai P.}
\REVIEW{Phys. Rev. Lett}{92}{2004}{224501}.

\bibitem{BF99}
\Name{Balkovsky E. \and Fouxon A.}
\REVIEW{Phys. Rev. E}{60}{1999}{4164}.

\bibitem{private} K.~Gaw\c{e}dzki, {\it Private communication}. 

\bibitem{K94}
\Name{Kraichnan R.H.}
\REVIEW{Phys. Rev. Lett.}{72}{1994}{1016}.

\end{thebibliography}
\end{document}